\documentclass[runningheads]{llncs}
\usepackage{graphicx}
\usepackage{makecell}
\usepackage{booktabs}

\usepackage{listings}
\usepackage{subcaption}
\begin{document}
\title{RAG-IT: Retrieval-Augmented Instruction Tuning for Automated Financial Analysis - A Case Study for the Semiconductor Sector}

\author{Hai-Thien To \and Tien-Cuong Bui \and Van-Duc Le}

\institute{
\begin{minipage}[t]{0.31\textwidth}
\centering
\textbf{Hai-Thien To} \\
University of Transport Technology \\
 Hanoi, Vietnam \\
\email{thienth@utt.edu.vn}
\end{minipage}
\hfill
\begin{minipage}[t]{0.31\textwidth}
\centering
\textbf{Tien-Cuong Bui} \\
Arontier Co., Ltd. \\
Seoul, South Korea \\
\email{cuongbt91@gmail.com}
\end{minipage}
\hfill
\begin{minipage}[t]{0.31\textwidth}
\centering
\textbf{Van-Duc Le} \\
Independent Researcher \\
Seoul, South Korea \\
\email{vanduc103@gmail.com}
\end{minipage}
}
%
\titlerunning{RAG-IT: Retrieval-Augmented Instruction Tuning LLMs}
%

\maketitle              
\begin{abstract}

Financial analysis relies heavily on the interpretation of earnings reports to assess company performance and guide decision-making. Traditional methods for generating such analyzes require significant financial expertise and are often time-consuming. With the rapid advancement of Large Language Models (LLMs), domain-specific adaptations have emerged for financial tasks such as sentiment analysis and entity recognition. This paper introduces RAG-IT (Retrieval-Augmented Instruction Tuning), a novel framework designed to automate the generation of earnings report analysis through an LLM fine-tuned specifically for the financial domain. Our approach integrates retrieval augmentation with instruction-based fine-tuning to enhance factual accuracy, contextual relevance, and domain adaptability. We construct a sector-specific financial instruction dataset derived from semiconductor industry documents to guide the LLM adaptation to specialized financial reasoning. Using NVIDIA, AMD, and Broadcom as representative companies, our case study demonstrates that RAG-IT substantially improves a general-purpose open-source LLM and achieves performance comparable to commercial systems like GPT-3.5 on financial report generation tasks. This research highlights the potential of retrieval-augmented instruction tuning to streamline and elevate financial analysis automation, advancing the broader field of intelligent financial reporting.

\keywords{LLMs \and financial analysis \and instruction tuning.}
\end{abstract}
\section{Introduction}
Large Language Models (LLMs) have shown remarkable proficiency in diverse applications, including financial analysis. BloombergGPT \cite{bloomberggpt}, a specialized LLM for finance, is built from scratch and excels in various financial tasks, though it is costly due to its use of proprietary data and training expenses of over \$2 million. Conversely, FinGPT \cite{fingpt} is an open-source low-cost LLM fine-tuned on instructional data, focusing on basic financial tasks like sentiment classification and named entity recognition. This paper explores more complex financial challenges, particularly earnings report analysis.

A company's earnings report serves as a comprehensive self-declaration regarding its financial performance within a given fiscal period. Consequently, earnings analysis assumes a pivotal role in financial assessment, aiming to garner a nuanced understanding of the company's performance. Therefore, earnings report analysis necessitates adept handling of various financial intricacies, including considerations such as the peer sector dynamics, challenges related to calendarization, and the nuanced language employed in the executives' communications. In the next section, we provide details about how earnings analysis can be conducted via our augmented system.

Instruction tuning, as proposed by \cite{alpaca}, stands out as a cost-effective and efficient technique for aligning an LLM with a specific downstream task, such as the earnings analysis. The process involves the collection of instruction-following data, subsequently used to fine-tune the LLM. Since human generation datasets are very costly, a pragmatic alternative involves initiating the fine-tuning process with seed examples and employing a high-capacity teacher LLM such as OpenAI GPT-3.5 to generate a diverse set of instructional data. While this approach has demonstrated success across various problem domains, such as general question-answering \cite{alpaca} and visual understanding \cite{llava}, its applicability to earnings report analysis remains an open question.

In addressing domain-specific challenges, an LLM frequently leverages the retrieval augmented generation (RAG) approach \cite{rag}, a methodology that provides the model with domain-specific context. In our work, aimed at grounding generation instructions within the financial domain, we introduce a novel method termed \textbf{retrieval-augmented instruction data generation}. This methodology integrates both the financial domain context and the instructional prompt during data collection, guiding the teacher LLM in generating questions and answers aligned with the provided context. The approach yields a rich set of financial instruction-following data, important for the fine-tuning of a financial-augmented LLM.

In summary, this research makes several significant contributions. Firstly, we address the earnings report analysis problem, representing a pivotal extension in the realm of LLMs focused on finance. Our novel approach involves a retrieval-augmented instruction data generation method designed for domain-specific tasks. This method generates instruction data contextually grounded in the financial domain. By fine-tuning an LLM with the curated instruction data, we demonstrate superior performance within the financial domain, much better than the well-known open-source LLM, Llama-2-7b \cite{llama2}, and comparable with the commercial GPT-3.5 model.

\section{Structured Components of Earnings Report Analysis}

A comprehensive earnings report analysis goes beyond the computation of financial metrics; it involves a structured reasoning process across several interconnected dimensions of company performance. To ensure both interpretability and reproducibility of the analysis, we decompose the process into a series of \emph{analysis modules}, each corresponding to a specific section of an earning report. These modules form the basis of our data-generation and report-synthesis framework.

Each instruction type encapsulates a distinct analytical objective, as detailed below. To ensure that our six-component structure is not arbitrary, we consulted two financial experts with MBA degrees and many years of experience in large corporations. Their feedback confirmed that this decomposition aligns well with the workflow used in professional earnings reports, providing practical grounding for our design choices.

\subsection*{(1) Company Core Information}
This section introduces the company's background, business model, and sectoral context. It provides foundational context for the subsequent financial and strategic assessments. Typical questions at this level include identifying the company's primary business domains, product lines, and target markets. The main information sources are the company's press releases and official websites, which ensure factual accuracy.  
This step grounds the report-generation process, enabling the model to contextualize later KPI-based interpretations within the company's operational scope.

\subsection*{(2) Key Financial Indicators (KPIs)}
The next stage involves extracting and interpreting \emph{core quantitative metrics} that define the company's financial performance. Examples include revenue, net income, operating margin, earnings per share (EPS), and cash flow. These KPIs are retrieved from earnings reports and press releases, then aligned with the analyst's custom-defined KPI set. This allows the analysis to be both standardized and flexible-standardized in structure, yet personalized to reflect the analyst's expertise.  
In our framework, these KPIs serve as \emph{anchor points} for the model's narrative generation, allowing the auto-generated analysis to maintain factual grounding while adapting to subjective analytical focus.  
Indeed, financial KPIs have been characterized as metrics that provide insights into underlying business performance across categories such as profitability, liquidity, efficiency, valuation and leverage. \cite{kpi1,kpi2}

\subsection*{(3) Comparative Analysis}
To provide market context, comparative analysis benchmarks the company's performance against peers in the same industry. This may involve ratio-based comparisons (e.g., revenue growth rate, profit margin) or relative ranking within the sector.  
The comparison step enhances interpretability by situating a company's performance within broader market trends. The relevant documents include both earnings reports of competitors and industry summaries, allowing the model to contextualize differences in business cycles or strategic focus.

\subsection*{(4) Outlook Analysis}
Forward-looking analysis assesses management's guidance, analyst consensus, and projected financial trends. By drawing from press releases, earnings reports, and occasionally analyst forecasts, the model can synthesize expectations for future performance, such as projected revenue growth or cost-optimization strategies.  
This component also reflects qualitative signals, such as management tone or sentiment extracted from the earnings call transcript, which complement the quantitative KPIs. Integrating these elements allows the model to generate a balanced view that combines data-driven projections with managerial perspectives.

\subsection*{(5) Summary and Executive Sentiment}
The summary component synthesizes key insights from both financial and textual sources, such as executives' remarks and press announcements. This step aims to distill the overall narrative conveyed during the earnings call, highlighting management's confidence levels, strategic priorities, or risk acknowledgments.  
We use this component to train the model to capture \emph{soft indicators} that are not directly observable in numerical data but are critical for evaluating corporate outlook and investor sentiment.

\subsection*{(6) Integrated Analysis Report}
Finally, all the preceding elements are integrated into a cohesive \emph{earnings report analysis}. This module combines objective KPI evaluation, comparative benchmarks, and qualitative outlook synthesis to generate a holistic view of the company's performance in the given quarter. The structure and tone are inspired by professional equity research reports, ensuring the generated output is aligned with financial reporting standards.  
The analyst remains in the loop by reviewing or updating the KPI set, thereby guiding the model's focus in subsequent iterations.

\section{Financial Instruction Data}
Developing a financially augmented large language model (LLM) requires a well-structured and context-rich dataset that reflects the language, reasoning patterns, and analytical conventions of financial reporting. To achieve this, we construct a two-tier dataset which serves as the foundation for instruction tuning and retrieval-augmented fine-tuning.

\subsection{General Financial Instruction Data}

\begin{table}[ht]
\caption{Instruction prompt to generate general financial instruction data and a data sample of generated instruction-following data. ${num\_questions\_per\_chunk}$ is the number of data samples we want to generate for a chunk of text.}
\begin{center}
\begin{tabular}[width=1.0\linewidth]{|p{12cm}|}
\hline
\textbf{Instruction:} You are an earnings report analyst. Your task is to ask ${num\_questions\_per\_chunk}$ questions to understand a company, its financial report, and its key financial performance. The questions should be diverse in nature across the document. Restrict the questions to the context of the information provided.\\
\hline
\textbf{Question:} What is the fiscal year-end date for NVIDIA Corporation?\\
\textbf{Answer:} The fiscal year end date for NVIDIA Corporation is January 28.\\
\hline
\end{tabular}
\label{tab:general_financial_data}
\end{center}
\end{table}

To fine-tune a financially augmented LLM, we first require a diverse set of \textit{general financial instruction data} that reflect real-world financial document structures and semantics. Such data are generated from the contextual content of company financial disclosures, including quarterly and annual reports, management commentary, and earnings call summaries.  

The data generation process begins by segmenting each financial document into discrete, semantically coherent text chunks. Each chunk represents a self-contained financial context that can independently support the generation of multiple instruction-following pairs. Examples of such contexts include a company's identity, its industry classification, and reported key financial metrics. For instance, our segmentation of NVIDIA's quarterly earnings report yielded chunks addressing queries such as:  
\begin{itemize}
    \item \textit{What is NVIDIA?}  
    \item \textit{What are the main industry sectors of NVIDIA?}  
    \item \textit{What is the revenue of NVIDIA for the second quarter of fiscal year 2024?}  
    \item \textit{What is the GAAP and Non-GAAP earnings per diluted share for the quarter?}  
    \item \textit{How many shares did NVIDIA repurchase during the second quarter of fiscal 2024, and for what amount?}
\end{itemize}

Each chunk thus acts as a contextual unit from which the model can learn to map document-grounded financial text to relevant natural language queries. Using a teacher LLM such as \texttt{GPT-3.5-turbo}, we automatically generate instruction–response pairs based on the textual content of each chunk. The teacher model is prompted to extract domain-specific insights, numerical facts, and relational patterns (e.g., quarter-over-quarter growth, year-over-year comparison, dividend schedules, or share repurchases).  

To ensure balanced coverage, we set the configuration parameter ${num\_questions\_per\_chunk}$ to 10. For a single quarterly report containing approximately 40 well-defined chunks, this setup yields around 400 financial instruction–response pairs. These pairs collectively form the \textit{general financial instruction dataset}, which serves as the base layer for fine-tuning the financially augmented LLM.  

The resulting dataset encompasses multiple dimensions of financial understanding, including:
\begin{enumerate}
    \item \textbf{Entity-level understanding:} Identifying the company, fiscal period, and report structure.  
    \item \textbf{Numerical reasoning:} Computing and interpreting KPIs such as revenue, operating income, and EPS.  
    \item \textbf{Temporal reasoning:} Comparing performance across quarters and fiscal years.  
    \item \textbf{Policy and governance insight:} Understanding board authorizations, dividend policies, and share repurchase programs.  
\end{enumerate}

This structured data generation process enables the LLM to capture the linguistic and analytical characteristics of financial narratives, providing a strong foundation for subsequent retrieval-augmented fine-tuning. An illustrative sample of the generated instruction-following data is shown in Table~\ref{tab:general_financial_data}.  

In summary, the general financial instruction dataset serves two purposes: (1) it provides a systematic foundation of financial question–answer pairs derived directly from authentic corporate disclosures, and (2) it establishes the contextual grounding necessary for extending the LLM's capacity to handle retrieval-augmented financial reasoning tasks discussed in the next section.


\subsection{Earnings analysis seed instructions}

\begin{table}[htbp]
\caption{Seed instructions for earnings analysis instruction-following data generation.}
\centering
\begin{tabular}{c|>{\centering\arraybackslash}m{2.5cm}|m{5.5cm}|m{3.0cm}}
\toprule
\textbf{No.} & \multicolumn{1}{c|}{\textbf{\thead{Instruction\\Type}}} & \multicolumn{1}{c|}{\textbf{Prompt Example}} & \multicolumn{1}{c}{\textbf{\thead{Relevant\\ Documents}}} \\
\midrule
1 & Company core information & What is Nvidia and its business sector? & Press Release \\
\hline
2 & Key financial indicators & What is the revenue? & Press Release, Earnings Report \\
\hline
3 & Comparison & Can you compare the revenue of Nvidia with its peer group in the semiconductors industry? & Press Release, Earnings Report \\
\hline
4 & Outlook & Can you give the outlook for the revenue of Nvidia in the next quarter? & Press Release, Earnings Report \\
\hline
5 & Summary & Can you summarize the earnings report and executives' statements of Nvidia? & Press Release, Earnings Call Transcript \\
\hline
6 & Analysis & Given the information above, can you generate an earnings report analysis for Nvidia in this quarter? & Equity Research Report\\
\bottomrule
\end{tabular}
\label{tab:seed_instruction}
\end{table}


Precise guidance for the teacher LLM in the form of seed instructions is essential to ensure the generation of accurate, diverse, and contextually grounded instruction-following data. Given the complexities inherent in financial text, particularly in earnings reports, the model must understand not only numerical facts but also qualitative relationships, comparative metrics, and managerial sentiment. To this end, we design a structured set of \textit{seed instructions} that guide the teacher model to produce high-quality financial reasoning data.

The seed instructions are crafted to emulate the reasoning process of human financial analysts. They cover multiple analytical dimensions, including factual extraction, comparison, temporal reasoning, and interpretive commentary. We employ conversation-based instruction prompts to create a natural flow of inquiry-mimicking how an analyst might sequentially explore a company's quarterly performance. This contextual chaining enables the teacher model to capture dependencies between questions and answers, resulting in coherent and analytically rich datasets.

Table~\ref{tab:seed_instruction} presents a taxonomy of six instruction types with representative examples and their associated document sources. Each instruction type plays a distinct role in the generation of financial instruction-following data:

\begin{itemize}
    \item \textbf{Company Core Information:} Focuses on the company's identity, business model, and operational domain. These questions establish the foundational context of the analysis.

    \item \textbf{Key Financial Indicators:} Targets quantitative performance metrics such as revenue, earnings per share (EPS), or operating margin. The responses typically include both the reported values and brief interpretations.

    \item \textbf{Comparison:} Involves benchmarking company metrics against peer firms or historical performance to capture relative strengths or weaknesses.

    \item \textbf{Outlook:} Encapsulates management guidance, analyst consensus, and forward-looking statements that indicate projected financial performance. These prompts encourage narrative synthesis beyond raw numbers.

    \item \textbf{Summary:} Integrates insights from multiple sections of the earnings release and call transcripts, summarizing the quarter's highlights and strategic focus.

    \item \textbf{Analysis:} Represents the highest level of reasoning, combining numerical interpretation, strategic evaluation, and investor sentiment to form a holistic analysis. This aligns most closely with the structure of an equity research report.
\end{itemize}

Each instruction type is linked to specific source documents—such as press releases, earnings reports, call transcripts, or equity research summaries—to ensure grounding in verified financial text. During instruction data generation, the teacher LLM is conditioned on both the prompt and its associated document snippet, allowing the resulting responses to be evidence-based and contextually aligned.

Furthermore, we design the prompting sequence to include interdependent instructions, where later prompts (e.g., \textit{Outlook} or \textit{Analysis}) depend on information extracted in earlier steps (e.g., \textit{Key Financial Indicators}). This multi-turn structure improves the model's capacity to integrate cross-referenced information, mirroring the analytical reasoning process of human experts.

The seed instructions thus serve as a blueprint for generating a comprehensive and multi-dimensional financial dataset. By training the student model on these structured instruction–response pairs, the LLM learns to:
\begin{enumerate}
    \item Accurately extract and interpret quantitative metrics from textual disclosures.
    \item Contextualize these metrics within industry and temporal frames.
    \item Generate coherent narratives that explain the causes and implications of financial outcomes.
\end{enumerate}

In summary, earnings analysis seed instructions establish a consistent and interpretable framework for generating high-quality financial reasoning data. They ensure that the model learns not only to retrieve factual content but also to perform structured analysis—an essential step toward developing a retrieval-augmented financial LLM capable of producing explainable and analyst-like insights.

\section{Retrieval-Augmented Instruction Tuning Method}
This section describes our retrieval-augmented instruction tuning method and outlines how we construct a financially augmented large language model (LLM) through fine-tuning using general financial instruction data and retrieval-augmented financial reasoning data. The overall framework aims to enhance an LLM's ability to reason about financial information in context, improving its performance on document-grounded financial tasks.

\begin{figure}[htbp]
\centerline{\includegraphics[width=1.0\linewidth]{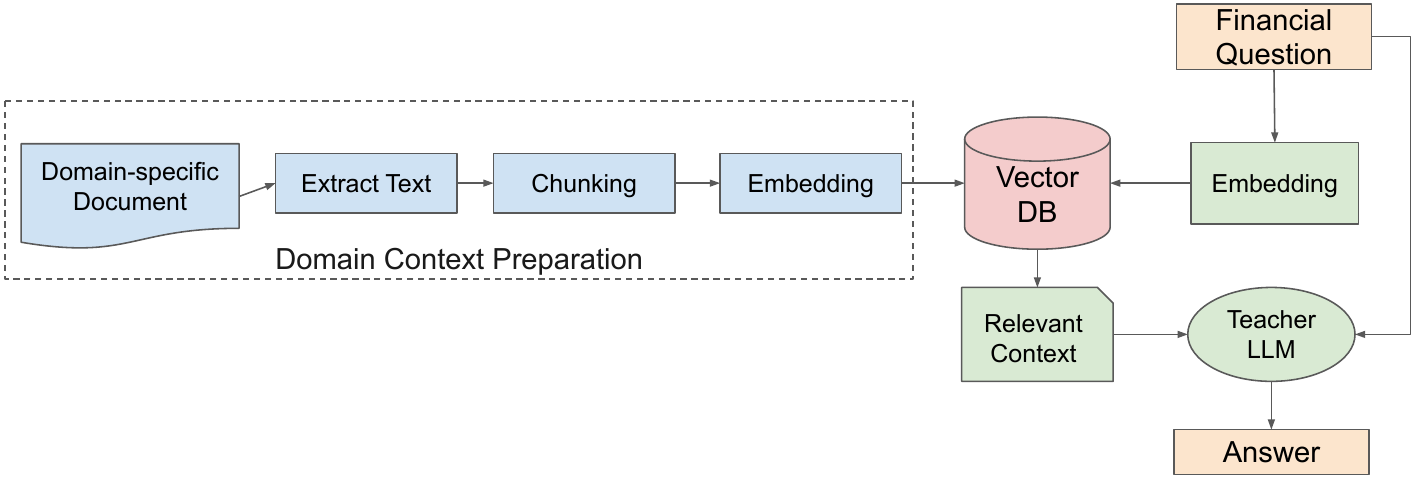}}
\caption{The retrieval-augmented instruction data generation workflow.}
\label{fig:rag_instruction_generation}
\end{figure}

In Fig.~\ref{fig:rag_instruction_generation}, we present the workflow of our retrieval-augmented instruction data generation and model fine-tuning methodology. The approach is designed to tightly couple context retrieval, instruction generation, and parameter-efficient fine-tuning. The method consists of two main components: (1) the retrieval-augmented instruction data generation process, and (2) the financially augmented LLM fine-tuning process.

\subsection{Retrieval-Augmented Instruction Data Generation}

This component focuses on creating context-grounded, instruction-following financial data derived from authentic financial documents. The goal is to generate diverse and accurate question–answer pairs that reflect the reasoning tasks encountered in financial analysis.

The pipeline begins with a \textbf{document indexing and retrieval} stage. Financial documents such as quarterly earnings reports, press releases, and SEC filings are segmented into coherent text chunks. These chunks are then embedded using a dense embedding model and stored in a vector database (e.g., ChromaDB), which supports semantic retrieval.

Given a financial topic or company name, the system retrieves relevant chunks based on similarity search. These chunks are passed as contextual input to a teacher model (e.g., \texttt{GPT-3.5-turbo}), which generates instruction–response pairs grounded in the retrieved financial context. Each generated sample follows the structure:
\begin{verbatim}
{
  "instruction": <financial query>,
  "input": <retrieved document context>,
  "output": <teacher-generated answer>
}
\end{verbatim}

This process ensures that generated data remain faithful to real financial statements and reduce hallucinations by grounding responses in retrieved evidence. In addition, conversational-style prompting is employed to elicit richer, more analytical responses that capture patterns such as revenue growth, expense variation, or year-over-year comparisons.

After generation, the data undergo normalization (e.g., numeric unit alignment, temporal standardization, duplicate removal). The final dataset integrates both general and analytic instructions, supporting a wide range of reasoning patterns that align with financial interpretation and decision-making.

\subsection{Financially Augmented LLM Fine-Tuning}

Once the retrieval-augmented financial instruction dataset is prepared, we fine-tune a pre-trained LLM to develop a financially specialized reasoning capability. The fine-tuning procedure focuses on efficiency and stability while preserving general language understanding.


During training, each instruction sample ($sample$) follows the previously described schema, where the instruction represents a financial query, the input provides relevant context, and the output serves as the correct answer. The model learns to interpret financial context, retrieve relevant details, and synthesize grounded analytical conclusions.

This approach ensures that the fine-tuned LLM can:
\begin{itemize}
    \item Perform context-grounded financial question answering with factual precision.
    \item Conduct multi-hop reasoning, such as temporal comparisons or cross-metric correlations.
    \item Generate human-like financial commentary aligned with analyst-style interpretations.
\end{itemize}

In summary, the retrieval-augmented instruction tuning framework integrates contextual data generation and parameter-efficient fine-tuning to build a domain-adapted financial LLM. The retrieval process ensures factual grounding, while instruction tuning refines the model's analytical expressiveness. Together, these two components yield a system capable of delivering accurate, interpretable, and contextually grounded financial reasoning.

\section{Experimental results}
\subsection{Experiment configurations and set up}
We employ \textbf{Llama-2-7B}~\cite{llama2} as the base model and fine-tune it using the \textbf{Low-Rank Adaptation (LoRA)} method~\cite{lora}. Following the QLoRA setup~\cite{qlora}, we enable 4-bit (NF4) quantization using \texttt{prepare\_model\_for\_kbit\_training}. Our LoRA configuration uses a rank of \textbf{64}, \textbf{lora\_alpha = 16}, and a dropout of \textbf{0.1}, applied to all attention and feed-forward projection layers with no bias adaptation. Fine-tuning is conducted using \texttt{SFTTrainer} with a maximum sequence length of \textbf{1024} tokens and packing enabled to improve training throughput.

We train for \textbf{3 epochs} with a global batch size of \textbf{8} (per-device batch size 2, gradient accumulation 2) using the \texttt{paged\_adamw\_32bit} optimizer. The learning rate is set to \textbf{2e-5} with a constant schedule and a warmup ratio of \textbf{0.03}. We disable gradient checkpointing and use \texttt{ddp\_find\_unused\_parameters=False} for efficient multi-GPU execution. The maximum gradient norm is clipped to \textbf{0.3}. 

We employed the Google Cloud virtual machine for training on four \textbf{Tesla T4 GPUs}, each equipped with 16GB of RAM. Due to the limited compute capability and memory bandwidth of T4 GPUs combined with 4-bit QLoRA dequantization overhead and activation packing, the end-to-end fine-tuning of our \textbf{800-sample} dataset requires approximately \textbf{40 hours}.

For the retrieval component, we use the \textbf{text-embedding-3-large} encoder to embed documents. The corpus is chunked into \textbf{512-token} segments with a \textbf{128-token} overlap. At inference time, the retriever selects the top \textbf{k=3} most relevant chunks using cosine similarity.

To ensure the correctness of GPT-3.5–generated training data, we apply a two-stage quality control process: (1) \textbf{rule-based filtering} (format validation, length constraints, and inconsistency checks), and (2) \textbf{manual inspection} of a randomly sampled \textbf{15\%} subset to identify hallucinations, numerical inconsistencies, and unclear instructions. Problematic items are revised or removed before final training.

We train our model on the financial instruction dataset generated on earnings reports from two companies in the semiconductor sector, Nvidia and AMD, in the third quarter of 2023 and evaluate the earnings report of Broadcom Inc. in the same period. In total, we have 800 training samples of financial instruction data.

\subsection{Evaluation metrics}
In this section, we present both quantitative and qualitative outcomes from our experimentation with a financial-augmented LLM using samples of earnings analysis question-answer (QA). To benchmark our model's performance, we compared it against two baselines: Llama-2, a general open-source LLM from Meta \cite{llama2}, and GPT-3.5, a commercial LLM from OpenAI. All three models were equipped with the same financial contexts and questions. Our evaluation leverages two key metrics: correctness and semantic similarity. Correctness assesses the model's ability to provide accurate answers to the given questions. As both the generated and ground-truth answers are in natural language, we define the evaluation scores on a scale from 1 to 10 and employ GPT-4 as the reference for judgment since GPT-4 can provide human-equivalent evaluation. Semantic similarity gauges how closely the generated responses align with the base answers, with lower values indicating better alignment.

The prompting for GPT-4 judgment is show in the listing \ref{listing:gpt4-prompt}.

\begin{lstlisting}[
breaklines=true,
    breakatwhitespace=true,
    columns=fullflexible,
caption={GPT-4 Judge Prompt}, 
label={listing:gpt4-prompt}
]
You are an expert evaluator. Your task is to compare a candidate answer 
against a reference answer and assign a numeric score based on the 
scoring rubric provided. You must follow the rubric strictly and output 
only the final score as an integer from 1 to 10.

Score 1: The answer is completely unrelated to the reference.
Score 3: The answer has minor relevance but does not align with the 
reference.
Score 5: The answer has moderate relevance but contains inaccuracies.
Score 7: The answer aligns with the reference but has minor errors.
Score 9: The numbers match perfectly, structure is similar.
Score 10: Completely accurate and perfectly aligned.
\end{lstlisting}

\subsection{Quantitative comparison}
In this section, we offer a quantitative analysis of our model in comparison to two baseline models, employing the aforementioned evaluation metrics of correctness and semantic similarity. The quantitative results are displayed in Table \ref{tab:quantative_compare}, unequivocally demonstrating the superior performance of our augmented model in the financial domain when contrasted with the open-source Llama-2. Remarkably, our model achieves comparable results to the commercial model, GPT-3.5. These findings underscore the efficacy and competitiveness of our approach, positioning it as a noteworthy advancement in the field.


\begin{table}[htbp]
\caption{Quantitative comparison of our Financially-augmented LLM with Llama-2-7b and GPT-3.5.}
\centering
\begin{tabular}{c|c|c}
\toprule
\textbf{Model} & \thead{\textbf{Correctness}\\\textbf{(1: worst, 10: best)}} & \thead{\textbf{Semantic Similarity}\\\textbf{(smaller is better)}} \\
\midrule
Financially-augmented LLM & 4.6 & 0.14427 \\
Llama-2-7b & 2.8 & 0.19126 \\
GPT-3.5 & 5.3 & 0.10659\\
\bottomrule
\end{tabular}
\label{tab:quantative_compare}
\end{table}

\subsection{Qualitative comparison}

\begin{table*}[ht!]
\centering
\caption{Qualitative comparison of model responses across different financial question types.}
\label{tab:qualitative_compare}

\begin{subtable}{0.95\textwidth}
\centering
\caption{Question Type: General Company Information}
\label{tab:qualitative_description}

\begin{tabular}{rl}
\textbf{Question:} & What is Broadcom Inc.? 
\vspace{4pt}
\end{tabular}

\begin{tabular}{|p{2.5cm}|p{9cm}|}
\hline
\textbf{\thead{Model}} & \textbf{\thead{Generated Answer}} \\
\hline
Our Financially-augmented LLM &
Broadcom Inc. is a technology company that designs, develops, and supplies
semiconductor and infrastructure software solutions. \\
\hline
Llama-2-7b &
1. An American manufacturer of semiconductor devices and infrastructure software solutions. 
2. A software company that provides solutions for managing applications. \\
\hline
GPT-3.5 &
Broadcom Inc. is a global technology leader that designs, develops, and supplies
semiconductor and infrastructure software solutions. \\
\hline
\end{tabular}
\end{subtable}

\vspace{0.4cm}

\begin{subtable}{0.95\textwidth}
\centering
\caption{Question Type: Financial Metric Extraction Query}
\label{tab:qualitative_numeric}

\begin{tabular}{rp{9cm}}
\textbf{Question:} & What is the company's quarterly dividend for the third quarter of 
Fiscal Year 2023? \\
\textbf{Ground-truth:} & \$4.60 per share. 
\vspace{4pt}
\end{tabular}

\begin{tabular}{|p{2.5cm}|p{9cm}|}
\hline
\textbf{\thead{Model}} & \textbf{\thead{Generated Answer}} \\
\hline
Our Financially-augmented LLM &
\$4.60 \\[2pt]

\hline
Llama-2-7b &
\$368.40 \\[2pt]
\hline
GPT-3.5 &
The company's quarterly dividend for the third quarter of Fiscal Year 2023 is \$4.60. \\[2pt]
\hline
\end{tabular}
\end{subtable}

\end{table*}


We conducted a rigorous qualitative comparison of our augmented model with two baseline models. To ensure a fair evaluation, all three models were provided with the same quarterly report document, and we employed an identical embedding model for vector indexing. Table~\ref{tab:qualitative_compare} presents examples of questions related to earnings analysis, along with the corresponding generated answers from each model. The qualitative analysis reveals clear distinctions in the quality and contextual relevance of responses.

In the general company information question, all three models, the open-source general-purpose Llama-2-7b, our Financially-augmented LLM, and the commercial GPT-3.5, produce correct and sufficiently detailed descriptions of Broadcom Inc. This indicates that broad corporate identity queries can be reliably answered even without financial-domain specialization. \textbf{However, in the financial metric extraction query, our model provides a more accurate and semantically aligned answer than the open-source baseline and shows a level of correctness that closely matches the response from the powerful commercial GPT-3.5 model.}

\paragraph{Our Financially-Augmented LLM.}
Our model consistently demonstrates high contextual accuracy and precise numerical reasoning. For instance, in the dividend question, the generated response (\textit{"4.60"}) directly matches the ground-truth value and preserves factual consistency with the source document. The model effectively retrieves and integrates the relevant context from the financial report, yielding concise and unambiguous answers. This indicates that the retrieval-augmented instruction tuning process allows the model to better internalize document structure and financial semantics. Moreover, it produces domain-appropriate terminology (e.g., “semiconductor and infrastructure software solutions”) comparable in style and factual fidelity to human financial analysts.

\paragraph{Llama-2-7b Baseline.}
The baseline Llama-2-7b model exhibits limited grounding and weaker numerical reasoning. Although it provides a generally correct company description, the responses show redundancy and lack of focus, as evidenced by the listing of multiple partial statements. For numerical queries, the model frequently produces spurious values (e.g., "368.40"), demonstrating its inability to extract and reason over financial figures from context. These errors highlight the lack of document-grounded alignment and domain-specific adaptation in the base LLM.

\paragraph{GPT-3.5 Teacher Model.}
GPT-3.5 generates highly coherent and semantically accurate responses, often matching or exceeding the clarity of human-written financial summaries. Its answer for the dividend question exactly reproduces the ground truth, confirming strong factual precision and contextual comprehension. However, GPT-3.5 relies heavily on its large-scale pretrained knowledge and less on explicit document retrieval, occasionally leading to generic phrasing or assumptions when context is ambiguous. Despite this, it serves as an effective teacher model for generating context-grounded instruction data due to its consistent reasoning quality.

\paragraph{Overall Comparison.}
Overall, our Financially-Augmented LLM achieves a balance between GPT-3.5's reasoning depth and Llama-2's fine-tuning adaptability. The model effectively integrates retrieval context to ensure factual correctness while maintaining natural linguistic fluency. Qualitatively, it shows a clear improvement in interpretability, numeric reliability, and contextual grounding compared to the baseline. This result substantiates the effectiveness of retrieval-augmented instruction tuning in enabling domain-specialized financial reasoning within smaller, efficient LLMs.

\subsection{More Success and Failure Cases}
We provide additional success and failure cases of our model across different question types in Table \ref{tab:failure_cases}, including different key financial indicators including dividend queries, revenue-related metrics, period identification, and segment-level breakdowns. These failure cases span multiple companies (Broadcom, NVIDIA, and AMD), illustrating that errors may occur not only in numeric extraction but also in semantic alignment with the ground-truth context. For each example, we list the question, the financial metric type, the model's generated answer, and the judgment from the GPT-4 judge model.

\begin{table*}[ht]
\centering
\caption{Representative success and failure cases of our model across various financial metric types.}
\label{tab:failure_cases}
\renewcommand{\arraystretch}{1.3}

\begin{tabular}{|p{1.8cm}|p{1.2cm}|p{5.1cm}|p{2.5cm}|c|}
\hline
\thead{\textbf{Question}} & 
\textbf{\thead{Metric \\ Type}} &
\textbf{\thead{Generated \\ Answer}} & 
\textbf{\thead{Judge Model \\ Reasoning}} & 
\textbf{\thead{Score}} \\ 
\hline

What is the company's quarterly dividend for the third quarter of Fiscal Year 2023? &
dividend &
The company did not declare or pay any dividends for the third quarter of Fiscal Year 2023. &
The model states that no dividend was declared, contradicting the ground truth that a dividend was paid. &
1 \\
\hline

What is the reported revenue for the third quarter of Fiscal Year 2023? &
revenue-related &
The reported revenue for the third quarter of Fiscal Year 2023 is \$13,507 million, up 101\% from the same period last year. &
The model provides revenue of \$13,507M, while the correct figure is \$8,876M, indicating a significant numeric error. &
5 \\
\hline

Can you provide a breakdown of the revenue by business segment? &
segment-level breakdowns &
Based on the provided context, AMD's revenue by business segment for the three months ended July 1, 2023 and June 25, 2022 are as follows: * Data Center: \$1,321 million in net revenue for Q2 2023 and \$1,486 million in Q2 2022. * Client: \$998 million in net revenue for Q2 2023 and \$2,152 million in Q2 2022. * Gaming: \$1,581 million in net revenue for Q2 2023 and \$1,655 million in Q2 2022. * Embedded: \$1,459 million in net revenue for Q2 2023 and \$1,257 million in Q2 2022. Note that these figures are based on management's judgment and allocation of expenses to the reportable segments, and may not necessarily reflect actual profitability by segment. &
The model fabricates a detailed AMD revenue breakdown even though the reference document does not provide such information. &
1 \\
\hline

What is the quarterly period ended date? &
period identification &
Based on the context provided, the quarterly period ended date is July 1, 2023. &
The model answers July 1, 2023, but the correct period end date is July 30, 2023. &
7 \\
\hline

What is the revenue of NVIDIA for second quarter fiscal year 2024? &
revenue-related &
NVIDIA's revenue for the second quarter of fiscal year 2024 was \$13.51 billion, up 101\% from the same period last year and up 88\% sequentially. &
The model correctly provides \$13.51B and correctly add the percentage increase. &
10 \\
\hline

\end{tabular}
\end{table*}


\section{Related Work}

Large Language Models (LLMs) have been increasingly applied to financial analysis, with early domain-specific models like BloombergGPT \cite{bloomberggpt} and FinGPT \cite{fingpt} demonstrating the value of financial adaptation. Recent studies extend this trend toward more sophisticated analytical tasks. Jadhav and Mirza \cite{jadhav2025equity} provide a comprehensive review of 84 studies on LLMs in equity markets, emphasizing their applications in forecasting and portfolio management. Regulatory perspectives, such as the ESMA–Alan Turing Institute report \cite{esma2025llm}, further highlight the importance of responsible and verifiable adoption in finance.

Instruction tuning has emerged as an efficient strategy for aligning LLMs with domain-specific objectives. Han et al. \cite{han2025survey} present a comprehensive survey of instruction tuning methods, emphasizing alignment-centric paradigms. This paper also shows that data quality and domain diversity critically influence downstream task performance. In finance, Hirano and Imajo \cite{hirano2024construction} propose building instruction-tuned LLMs without explicit instruction datasets via continual pretraining and model merging, and Tanabe et al. \cite{tanabe2024jafin} introduce JaFIn, a Japanese financial instruction dataset that enhances domain adaptability.

Retrieval-Augmented Generation (RAG) further improves factual grounding and contextual accuracy in financial reasoning. Zeng \cite{zeng2025quantmcp} presents QuantMCP, a retrieval-grounded framework linking LLMs to verified financial data sources, while Wang et al. \cite{wang2025financial} evaluate RAG and fine-tuning approaches for financial-statement analysis, highlighting their complementary strengths. Despite these advances, the integration of retrieval augmentation and instruction tuning for automated earnings report analysis remains unexplored. This work introduces RAG-IT, a framework unifying retrieval-augmented instruction data generation and fine-tuning to address this gap in financial LLM research.

\section{Conclusion}
This research paper aims to apply the field of large language models to the financial domain, with a particular focus on addressing the challenges of analyzing earnings reports. The introduction underscores the significance of this extension for researchers investigating the capabilities of large language models in the financial sector. The paper introduces an innovative approach centered around a retrieval-augmented instruction data generation method tailored specifically for tasks within the financial domain. The results highlight the efficacy of the augmented model, showcasing its potential to democratize the process of earnings report analysis.

Nevertheless, our approach has several limitations. RAG-IT may struggle with highly nuanced questions that require deep domain expertise, multi-hop reasoning over various financial metrics, or precise numerical computation. Its accuracy also depends heavily on the completeness and quality of retrieved evidences; missing or poorly retrieved context can degrade performance. Inference requires running both retrieval and generation, which introduces additional latency and computational cost compared to standard LLM inference. These limitations highlight important directions for future research, including more robust retrieval methods, improved reasoning capabilities, and more efficient architectures for financial analysis.

\bibliographystyle{splncs04}
\bibliography{bibliography}

\end{document}